\documentclass[pre,aps,floatfix,twocolumn,superscriptaddress,10pt]{revtex4-2}

\usepackage{amsmath,amssymb}
\usepackage{graphics,graphicx}
\usepackage{caption}
\usepackage{subcaption}
\usepackage{amsmath}
\usepackage{lineno}
\usepackage{hyperref}
\hypersetup{colorlinks=true, pdfstartview=FitV, linkcolor=blue, citecolor=black, plainpages=false,  urlcolor=blue}
\usepackage{cleveref}

\crefname{equation}{Eq.}{Eqs.}
\crefname{figure}{Fig.}{Figs.}

\begin{document}

\title{Acoustic emission data based modelling of fracture of glassy polymer}

\author{Subrat Senapati}
\email{subrat.senapati52@gmail.com}
\affiliation{Department of Applied Mechanics, Indian Institute of Technology Madras, Chennai-600036, India}
\author{Anuradha Banerjee}
\email{anuban@iitm.ac.in}
\affiliation{Department of Applied Mechanics, Indian Institute of Technology Madras, Chennai-600036, India}
\author{R.Rajesh} 
\email{rrajesh@imsc.res.in}
\affiliation{The Institute of Mathematical Sciences, C.I.T. Campus, Taramani, Chennai-600113, India} 
\affiliation{Homi Bhabha National Institute, Training School Complex, Anushakti Nagar, Mumbai-400094, India}

\date{\today}

\begin{abstract}
Acoustic emission (AE) activity data resulting from the fracture processes of brittle materials is valuable real time information regarding the evolving state of damage in the material. Here, through a combined experimental and computational study we explore the possibility of utilising the statistical signatures of AE activity data for characterisation of disorder parameter in simulation of tensile fracture of epoxy based polymer. For simulations we use a square random spring network model with quasi-brittle spring behaviour and a normally distributed failure strain threshold. We show that the disorder characteristics while have marginal effect on the power law exponent of the avalanche size distribution, are strongly correlated with the waiting time interval between consecutive record breaking avalanches as well as the total number of records. This sensitivity to disorder is exploited in estimating the disorder parameter suitable for the experiments on tensile failure of epoxy based polymer. The disorder parameter is estimated assuming equivalence between the amplitude distribution of AE data and avalanche size distribution of the simulations. The chosen disorder parameter is shown to well reproduce the failure characteristics in terms of the peak load of the macroscopic response, the power-law behaviour with avalanche dominated fracture type as well as realistic fracture paths. 

\end{abstract}

\maketitle

\section{Introduction}
\label{intro}
Brittle materials, such as concrete, wood, bone, coal etc., have inherent disorder in the form of micro-cracks, crystal defects, micro-voids, inclusions, micro-structural variation etc. Under mechanical stress, the inherent disorder often results in random independent nucleation of damage at multiple sites that subsequently exhibit correlated intermittent growth leading to rapid final failure. Acoustic
signals, elastic stress waves emitted by fracture processes, are valuable real-time information about the evolution of damage within the material from its initial state to its final collapse or failure~\cite{han2022,hu2022,guo2021,van2012}. While this information has been qualitatively widely utilised in identification of type of fracture mechanisms, their onset and progression, in both brittle as well as ductile materials,~\cite{lin2019,salminen2002,wang2023,botvina2023} the quantitative analyses of acoustic emission (AE) data have focussed primarily on establishing the existence of scale free behaviour and the associated power-law statistics~\cite{zhao2019,salje2018,planes2013,maes1998}. Predictive indicators from analyses of experimental AE data are vital in life assessment of existing structures~\cite{jiang2017,pal2016,danku2013}, however, utilising AE data for design and development of robust structural systems requires predictive fracture models that incorporate disorder by utilising AE data in characterisation of its model parameters.

 Stochastic discrete fracture models such as fibre bundle model~\cite{daniels1945,hansen2015,pradhan2010}, random fuse model~\cite{de1985random,zapperi1997,zapperi1999}, beam bending model~\cite{schlangen1997,raischel2005,sagar2019,van2002}, random spring network model (RSNM)~\cite{ray2006,kale2014,zapperi1997first} and other discrete element models~\cite{pal2016,kun2013,kun2014} have been shown to be effective in reproducing several statistical signatures observed in AE data of fracture experiments. Disorder in these models has been typically included by taking threshold for failure to be statistically distributed~\cite{herrmann1989,sahimi1986,senapati2023,curtin1990}, wider the distribution, higher the disorder. The avalanche of damage observed in the models, follows scale free power law statistics, as observed in experimental AE data~\cite{ray2006,zapperi1999,nukala2005,zapperi2005}. It has also been shown that the disorder plays a significant role in the type of fracture: nucleation type for low disorder, avalanche type for moderate disorder and percolation type in the limit of high disorder~\cite{senapati2023,shekhawat2013,kumar2022}. Apart from statistical signatures, discrete models have also shown to reproduce experimental macroscopic response as well as realistic fracture paths~\cite{kumar2022,boyina2015,mayya2018}, however, the disorder parameter was assumed on the basis of comparison with features of the macroscopic stress-strain response. 

More recently, AE data has also been analysed with the objective of finding statistical signatures of imminent collapse or final failure. The power-law exponent characterising the avalanche distribution has been shown to have a crossover between off-critical and near-critical regime both in experiments~\cite{jiang2017,jiang2016} as well as simulations based on fibre bundle model~\cite{pradhan2005}, random fuse model~\cite{pradhan2005}, a discrete element model of sedimenting spheres~\cite{pal2016}, as well as spring network model~\cite{senapati2023record}. The crossover in exponents has been seen to correspond to the change in the spatial location of AE events at different stages of deformation~\cite{zhao2019,salje2018,salje2021,iturrioz2014}. Initial stages have randomly distributed independent AE event sources that result in larger exponent in the size distribution. In the approach to final failure, the location of AE sources are concentrated in a band or process zone and are spatially correlated, the associated avalanche distribution exponent is lower in comparison~\cite{jiang2016}. Indications of imminent failure are also evident in the statistics of record breaking avalanches, record becomes those avalanches that have the highest amplitude compared to all previous avalanches of the time series. Record breaking avalanches form a subset of events of the complete time series of the avalanche size. Statistics of record breaking avalanches reveal that near final failure the record avalanches occur rapidly. The resulting reduction in waiting time interval between two consecutive record breaking events is a strong indicator  of imminent collapse~\cite{jiang2017,jiang2016,pal2016,kadar2022,danku2014,Roy2023,senapati2023record}. While real time AE data is useful in life assessment and prediction of failure of an existing structure, it has not been utilised as an input for predictive fracture modelling in the design and development stage of structures. 

In tensile fracture simulations of a representative material, is it possible to estimate the disorder characteristics by utilising the statistical signatures of AE data? In the present work, we address this question through a combined computational and experimental study of the fracture of an epoxy-based polymer. We perform fracture simulations using a square random spring network model where the spring behaviour is taken to be quasi-brittle, with linear and non-linear regimes, and a normally distributed failure strain threshold. Through a  parametric study, we identify the sensitivity of the power law exponent of the avalanche size distribution, the waiting time interval between consecutive record avalanches, as well as the total number of records to the disorder characteristics of the springs. This sensitivity to disorder is then exploited in estimating the disorder parameter suitable for reproducing experimental data from the tensile failure of epoxy-based polymer assuming equivalence between the amplitude distribution of AE data and avalanche size distribution of the simulations. The effectiveness of the chosen disorder parameter is demonstrated by showing the simulations reproduce the experimental failure characteristics in terms of the peak load of the macroscopic response, the power-law behaviour with avalanche-dominated fracture type, and realistic fracture paths. 

\section{Experiment}
\label{sec:2}

\subsection{Specimen preparation}
\label{sec:2.1}

\begin{figure}
\centering
\includegraphics[width=0.4\textwidth]{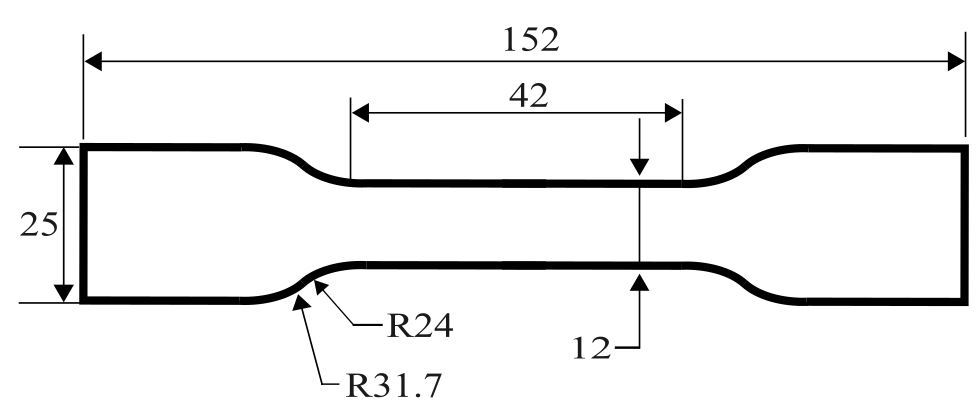}
\caption{The dimensions (in mm) of the mould used for preparing the dog-bone specimen.}
\label{fig31}
\end{figure}

\begin{figure}
\centering
\begin{subfigure}[b]{0.237\textwidth}
\centering
\includegraphics[width=3.4cm]{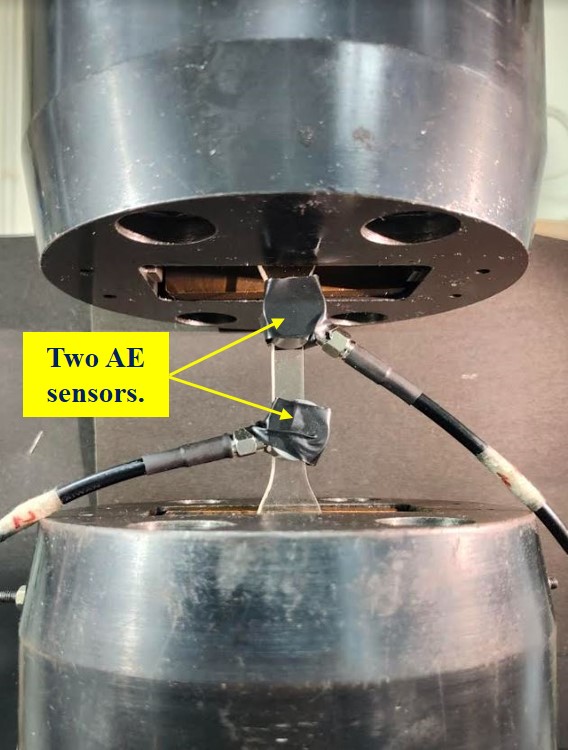}
\caption{}
\label{fig32a}
\end{subfigure}
\hfill
\begin{subfigure}[b]{0.237\textwidth}
\centering
\includegraphics[width=4.5cm, height=4.5cm]{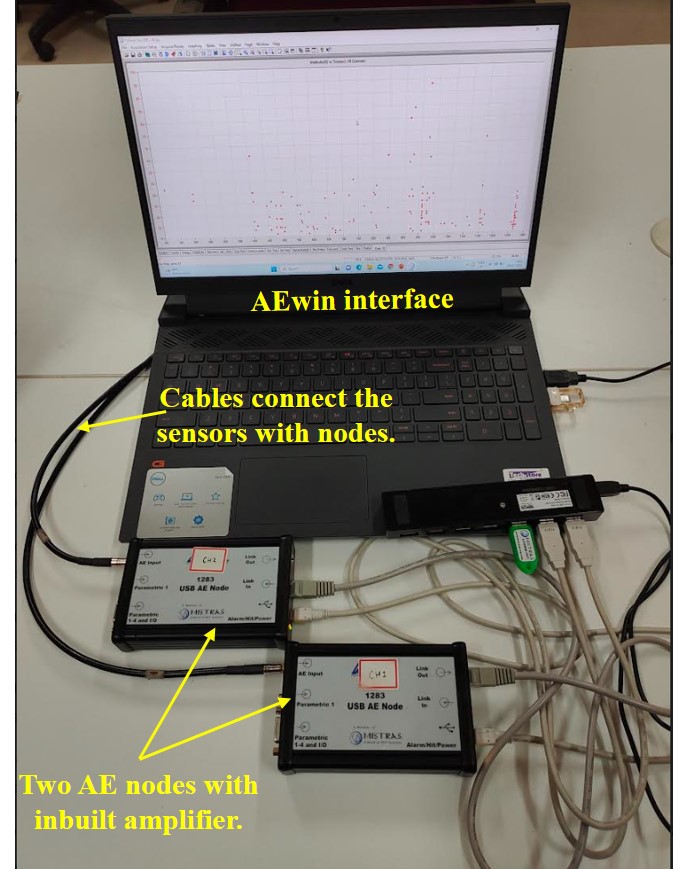}
\caption{}
\label{fig32b}
\end{subfigure}
\caption{Experimental setup showing (a) the sensors attached to the gauge section of the specimen and (b) the AE acquisition apparatus.}
\label{fig32}
\end{figure}

Dog-bone specimens are prepared from epoxy based resins using silicon rubber moulds with dimensions as shown in \cref{fig31}. Initially, the LY556 (Bisphenol A Diglycidyl Ether, DGEBA) resin and the HY951 (Triethylene tetraamine, TETA) curing agent are poured in a beaker at a ratio of 10:1 by weight. The solution is mixed for eight minutes using hand mixing method followed by a magnetic stirrer and then allowed to settle for two minutes. The epoxy solution is slowly poured into the mould, which has an inbuilt riser to compensate for shrinkage. The filled epoxy is cured in two stages; in the pre-curing stage, the mould is kept at room temperature for 3 hours, followed by curing in an oven at $60^\circ$C for 8 hours. The cured specimens are polished with 800 and 1000-grit grade sandpapers. The specimens are moulded in three batches [B1, B3, B4]. Finally, aluminium tabs are pasted on the upper and lower portions of the specimen to avoid local crushing at the grippet regions. 

\subsection{Mechanical testing}
\label{sec:2.2}
The dog-bone specimens are uniaxially tested by a servo-hydraulic type fatigue testing machine (DMG group) in displacement control mode at a rate of 0.1 mm/min. The crackling signals emitted by the specimens during the tensile test are captured by two piezoelectric sensors, as shown in \cref{fig32}. We apply grease on the sensor surface to maintain good acoustic coupling between the surfaces of the sensor and specimen. The sensors (R15$\alpha$), manufactured by Physical Acoustic Corporation (PAC), USA,  operate between  50-400 kHz and at a resonant frequency of 150 kHz. Two 1283 AE nodes (frequency responses between 1kHz to 1MHz) acquire the AE signals in real-time, and an inbuilt pre-amplifier amplifies the signals by 40 dB. The amplitude of these signals is computed in volts using the AEwin software. Before starting the test, the surface coupling between the sensor and specimen surface is checked and the attenuation in AE signals is minimised by repeated pencil lead break tests~\cite{ASTME976_15}. The noises from the surroundings and the machine set-up are filtered by setting a threshold of 31 dB. 

\begin{figure}
\centering
\begin{subfigure}[b]{0.237\textwidth}
\centering
\includegraphics[width=4.5cm]{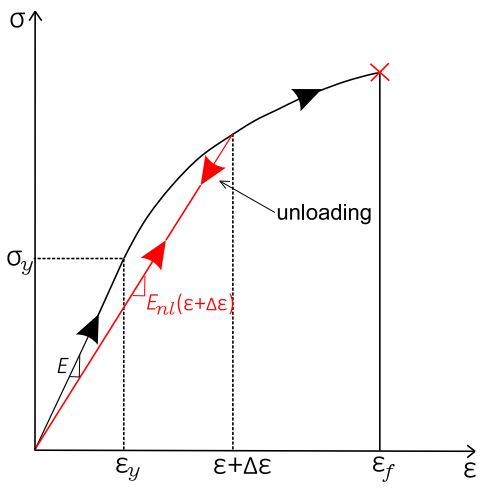}
\caption{}
\label{fig20a}
\end{subfigure}
\hfill
\begin{subfigure}[b]{0.237\textwidth}
\centering
\includegraphics[width=4.5cm]{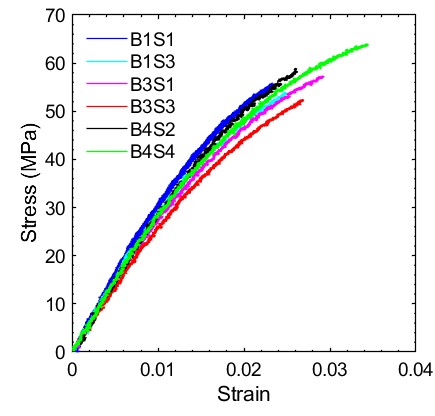}
\caption{}
\label{fig20b}
\end{subfigure}
\caption{(a) Response of a typical quasi-brittle material, and  (b) stress-strain response from the experiments.}
\label{fig20}
\end{figure}

During the experiment, the stress-strain data and the time-synced AE parameters are collected from the fatigue machine and AEwin software, respectively. A typical quasi-brittle stress-strain behaviour of such materials is shown in \cref{fig20a}, and the actual stress-strain responses of the specimens are in  \cref{fig20b}. The deviations from linear elastic response of the specimens are well within the range reported in the literature~\cite{Almeida1998}, the average strength of the specimens is approximately $57$ MPa and the observed variations in the strengths are expected due to variations in inherent defects across the specimens.  

\begin{figure}
\centering
\includegraphics[width=0.47\textwidth]{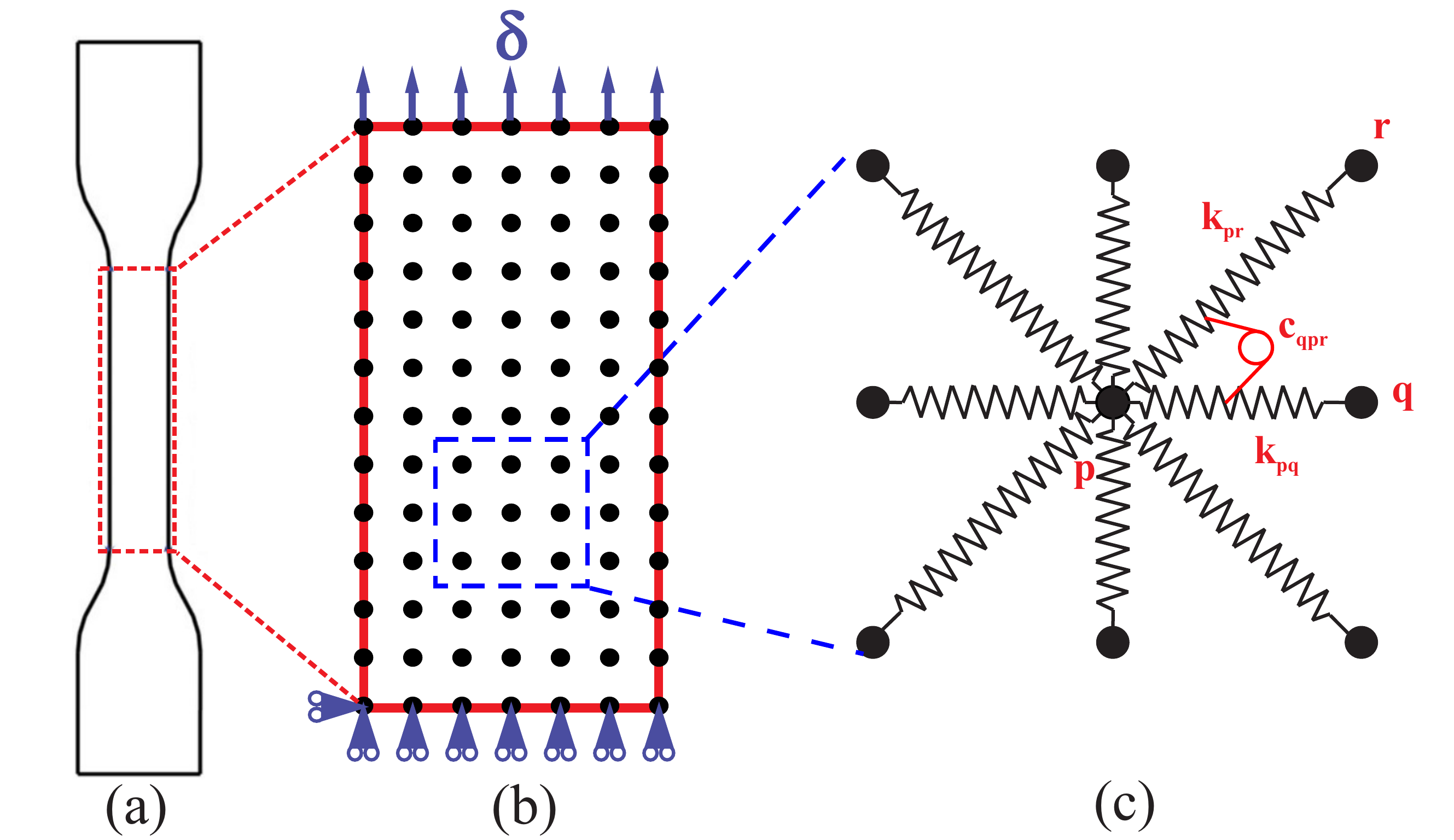}
\caption{(a) Schematic representation of a dog-bone specimen, (b) gauge section of the specimen is discretised with a square lattice spring network, and (c) the connectivity with neighbours of each lattice point.}
\label{fig21}
\end{figure}

\section{Model and details of simulation}
\label{sec:3}
In this section, we describe the structure and working principles of the random spring network model, followed by the characterisation of quasi-brittle behaviour exhibited by the material. Finally, we will estimate the model parameters for reproducing the constitutive behaviour and failure pattern of the epoxy-based glassy polymers.
\subsection{Model description}
\label{sec:3.1}
 In RSNM, we discretise the domain as highlighted in \cref{fig21}(a) with a square lattice, as shown in \cref{fig21}(b). Each lattice point ($p$) in the network is connected with its nearest neighbours ($q$) and diagonal neighbours ($r$) with extensional springs. The lattice point also interacts with a set of two adjacent neighbours by torsional spring (\cref{fig21}(c)). 

The spring constants are determined by equating the strain energy density of the continuum with the potential energy density of the spring network~\cite{monette1994}. The potential energy $\mathrm{\Phi}$ stored in the spring network due to elastic deformation  has contributions from both the extensional $\mathrm{\Phi_{ext}}$ and torsional $\mathrm{\Phi_{rot}}$ springs:     
\begin{equation}
\label{eq1}
\mathrm{\Phi = \Phi_{ext}+\Phi_{rot}}.
\end{equation}
The net potential energy stored accounting for the eight extensional springs of every lattice site is
\begin{equation}
\label{eq2}
\mathrm{\Phi_{ext}} = \sum_{\left\langle ij\right\rangle}\frac{1}{2}k_{ij}\left(|\vec{r}_{i}-\vec{r}_{j}|-a_{ij}\right)^2,
\end{equation} 
where $\vec{r}_{i}$ and $\vec{r}_{j}$ are the position vectors of points $i$, and $j$, $\left\langle \right\rangle$ denotes all pairs connected by extension springs and $a_{ij}$, $k_{ij}$ are the undeformed length and stiffness of the extensional spring connecting them. The net potential energy stored in the torsional springs is
\begin{equation}
\label{eq3}
\mathrm{\Phi_{rot}} = \sum_{\left\langle qpr\right\rangle}\frac{1}{2}c_{qpr}\left(\theta_{qpr}-\frac{\pi}{4}\right)^2,  
\end{equation}
where $c_{qpr}$ is the stiffness of the torsional spring and for an undeformed lattice, the angle subtended between $q$,$p$ and $r$, $\theta_{qpr}=\frac{\pi}{4}$. The summation $\langle qpr \rangle$ is over all possible combinations of the triads in the lattice. The potential energy density $\mathrm{\Phi}$ of the lattice is equated with the strain energy density of the continuum to evaluate the elastic parameters, Young's modulus $E$ and Poisson's ratio $\nu$ in terms of lattice parameters as~\cite{monette1994}:

 \begin{eqnarray}
 \label{eq4}
 E = \frac{8k\left(k+\frac{c}{a^2}\right)}{3k+\frac{c}{a^2}}, \\
 \label{eq5}
 \nu = \frac{\left(k-\frac{c}{a^2}\right)}{3k+\frac{c}{a^2}}, 
 \end{eqnarray}
where $c = c_{qpr}$, and $k$ is the stiffness of diagonal springs. The resulting stiffness of both vertical and horizontal springs would be $2k$. 

When the network is deformed, the positions of the particles are updated using Newton's equation of motion to achieve static equilibrium. The resultant force on a particle $p$ due to its interactions with its neighbours is given by  
\begin{equation}
\label{eq6} 
\vec{a}_p= -\nabla_{\vec{r}_p}\phi,
\end{equation}
where mass is set to unity. To achive equilibrium as well as to dampen excess oscillations, an extra dissipative term $-\gamma\vec{v}_p$ is also added. Once $\vec{a}_p$ is known, the updated position vector of the lattice point $p$ at time $t + \delta t$ is computed by the Verlet algorithm, using the last two position vectors $\vec{r}_p(t)$, $\vec{r}_p(t - \delta t)$ and $\vec{a}_p$ as:
\begin{equation}
\label{eq7} 
\vec{r}_p(t + \delta t) = \vec{r}_p(t)\left(2-\gamma\delta t\right)-\vec{r}_p(t - \delta t)\left(1-\gamma\delta t\right)+ \vec{a}_p (\delta t)^2,
\end{equation}
where the $\vec{v}_p$ is evaluated using the backward difference formulae $\vec{v}_p(t)=(r(t)-r(t-\delta t))/\delta t +O(\delta t)$. Within a loading step, Eq.$\left(\ref{eq7}\right)$ is iterated till the net kinetic energy of the system goes below a predefined threshold. Equilibrium is re-verified by ensuring that the magnitude of the net force at the top and bottom rows are equal within numerical error. If any spring breaks at the end of the equilibrium step, then the system is re-equilibrated. This iterative process continues within the loading step until no more springs break for the applied increment in strain. 

\subsection{Modelling the constitutive behaviour}
\label{sec:3.2}

The constitutive behaviour of the epoxy based polymer is quasi-brittle and has two distinct regimes, as shown earlier in \cref{fig20b}. In the model we take the response to be linear elastic upto the elastic limit, $\sigma_y$, and beyond that the response is taken to be quasi-brittle. The response of the material is modelled assuming a power-law strain hardening constitutive relation,
\begin{equation}
\begin{split}
\label{eq8}
\epsilon &= \frac{\sigma}{E}, \sigma\leq\sigma_y,\\
 &= \frac{\sigma_y}{E}\left(\frac{\sigma}{\sigma_y}\right)^{\frac{1}{n}},  \sigma>\sigma_y,
\end{split} 
\end{equation}
where $E$ is the elastic modulus, $\sigma_y$ is the elastic limit and $n$ is the power-law hardening exponent. For the spring behaviour, in the linear regime $(\epsilon\leq\epsilon_y)$, the spring constants are directly calculated from $E$ and $\nu$, using Eq.$\left(\ref{eq4},\ref{eq5}\right)$. But when the strain in a spring crosses $\epsilon_y$, firstly, the secant modulus $E_{nl}$ corresponding to the strain $\epsilon+\delta\epsilon$ is evaluated as~\cite{kumar2021}:
\begin{equation}
\label{eq9}
E_{nl}\left(\epsilon+\delta\epsilon\right) = E \left(\frac{\epsilon+\delta\epsilon}{\epsilon_y}\right)^{n-1}.
\end{equation}
While unloading, to avoid any changes to the secant modulus, it is updated as:
\begin{equation}
\label{eq10}
E\left(\epsilon+\delta\epsilon\right) = \mathrm{min}[E_{nl}\left(\epsilon+\delta\epsilon\right),E\left(\epsilon\right)].
\end{equation}
Finally, the spring constants in the nonlinear regime are calculated from the updated secant modulus $E$ and $\nu$ as per Eq.$\left(\ref{eq4},\ref{eq5}\right)$~\cite{kumar2021}.

In the model, a spring fails when it stretches beyond a threshold strain, and simultaneously, the rotational springs associated with the broken spring are also taken to be broken. The failure strain of the springs, $\epsilon_f$,  is assigned from a normal distribution, and its probability density function $P\left(\epsilon_f\right)$ is defined as
\begin{equation}
\label{eq11}
P\left(\epsilon_f\right) = \frac{1}{\sqrt{2\pi(\Delta\overline{\epsilon}_f)^2}}e^{-\frac{\left(\epsilon_f-\overline{\epsilon}_f\right)^2}{2(\Delta\overline{\epsilon}_f)^2}},
\end{equation}
where $\overline{\epsilon}_f$ is the mean and $\Delta\overline{\epsilon}_f$ is the standard deviation of the distribution, where $\Delta$ is a multiplicative factor, controlling extent of disorder.

\subsection{The model parameters}
\label{sec:3.3}

The highlighted gauge section of the dog-bone specimen \cref{fig21}(a) of size approximately $\mathrm{42\times10}$ $\mathrm{mm^2}$ is discretized with $NY\times NX$ lattice points having lattice spacing $a$. Hence, for the lattice spacings of a = 0.1, 0.2 and 0.4 mm, the $NY\times NX$ are $420\times 102$, $210\times 52$ and $105\times 26$, respectively. For simulating uniaxial tension, quasi-static tensile load is applied at a rate of $0.000436$ mm/step, as shown in \cref{fig21}(b). The boundary conditions of the loading configuration are:  vertically upward displacement is applied to the top row, vertical motion of the bottom row is restricted, and both the horizontal and vertical motion of the left bottom corner point is restricted. The material parameters, $E$ = 3.12 GPa, $\nu$ = 0.26, $n$ = 0.71, and $\epsilon_y$ = 0.0055, used for simulating the constitutive behaviour are calculated by fitting Eq.$\left(\ref{eq8}\right)$ to the stress-strain data for the B1S3 specimen. The damping constant $\gamma$ = 0.5 was found to suitably dampen the excessive oscillations of the system. In the following parametric study, we will only vary $a$, $\overline{\epsilon}_f$, and  $\Delta$, keeping the remaining parameters and geometry unchanged.   

\section{Results and Discussion}
\label{sec:4}

We begin with a computational parametric study on the effect of disorder on the fracture response, avalanche distribution and record statistics of quasi-brittle material. The parametric study will assist us in selecting an optimised degree of disorder suitable for modelling the glassy epoxy based polymer. Finally, the macroscopic responses, such as the elastic response, damage pattern, final breakage behaviour, and damage rate, will be compared with the corresponding experimental results to benchmark the effectiveness of the methodology. 

\subsection{Parametric study}
\label{sec:4.1}

\begin{figure}
\centering
\begin{subfigure}[b]{0.237\textwidth}
\centering
\includegraphics[width=4.5cm]{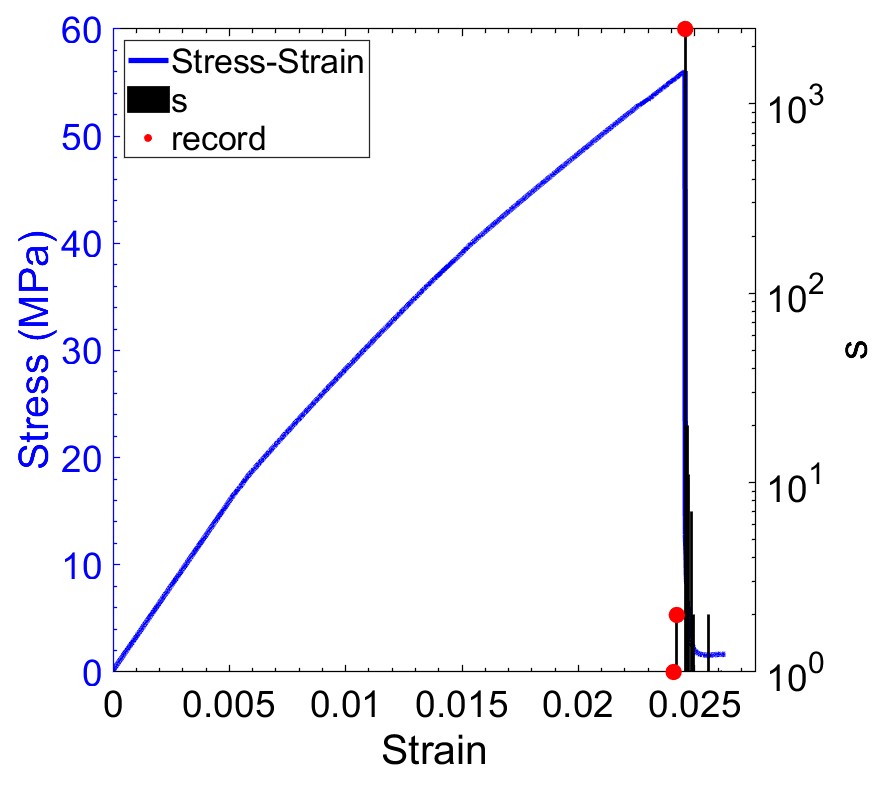}
\caption{}
\label{fig1a}
\end{subfigure}
\hfill
\begin{subfigure}[b]{0.237\textwidth}
\centering
\includegraphics[width=4.5cm]{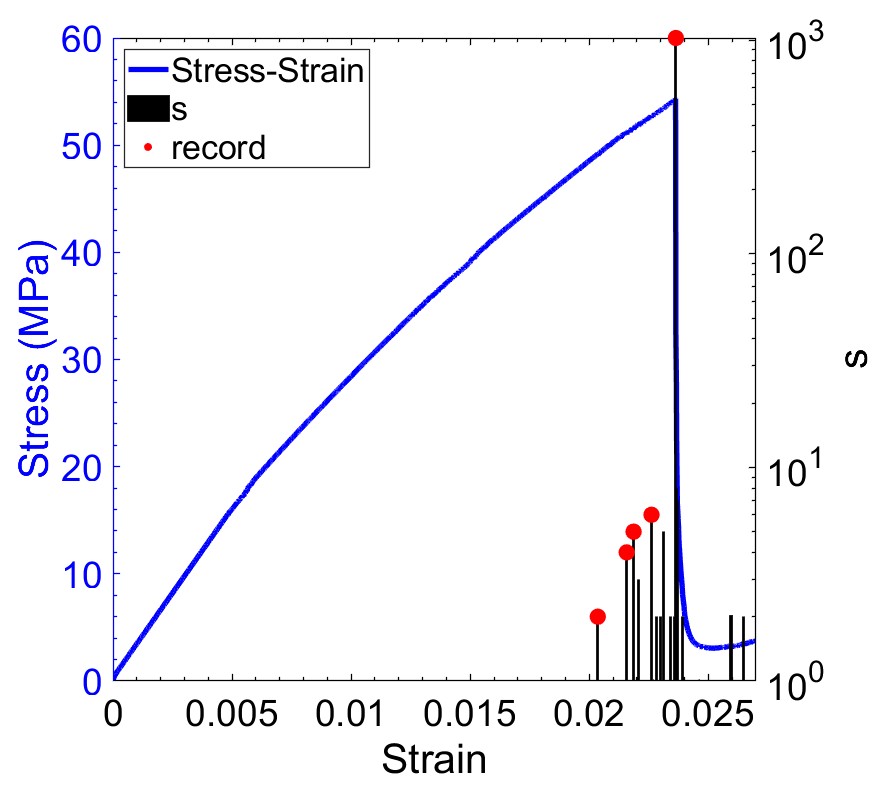}
\caption{}
\label{fig1b}
\end{subfigure}
\begin{subfigure}[b]{0.237\textwidth}
\centering
\includegraphics[width=4.5cm]{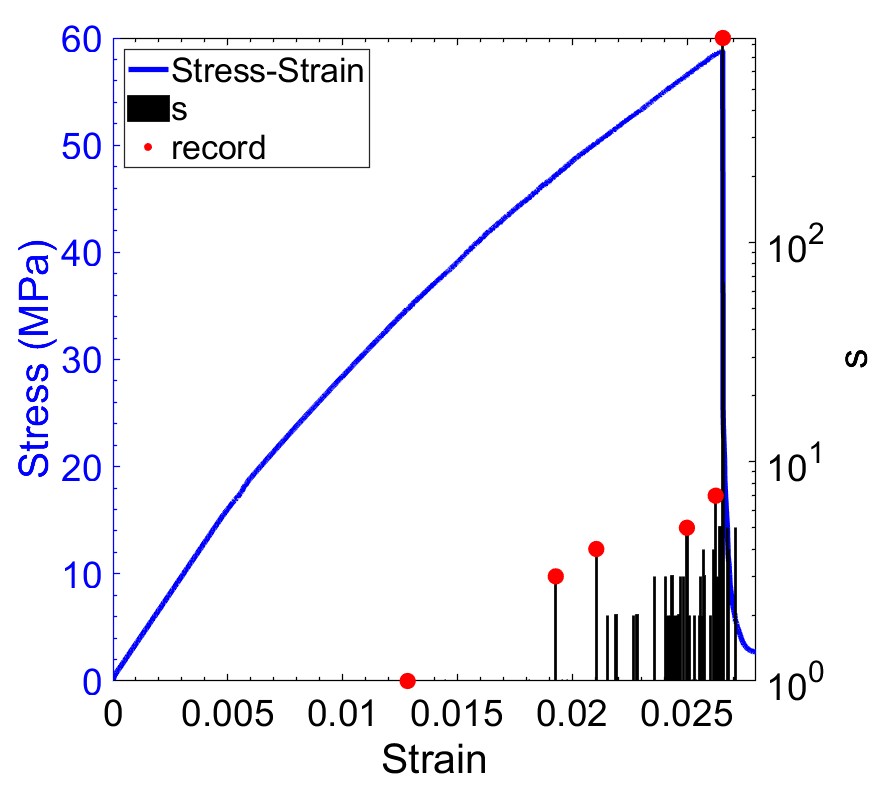}
\caption{}
\label{fig1c}
\end{subfigure}
\hfill
\begin{subfigure}[b]{0.237\textwidth}
\centering
\includegraphics[width=4.5cm]{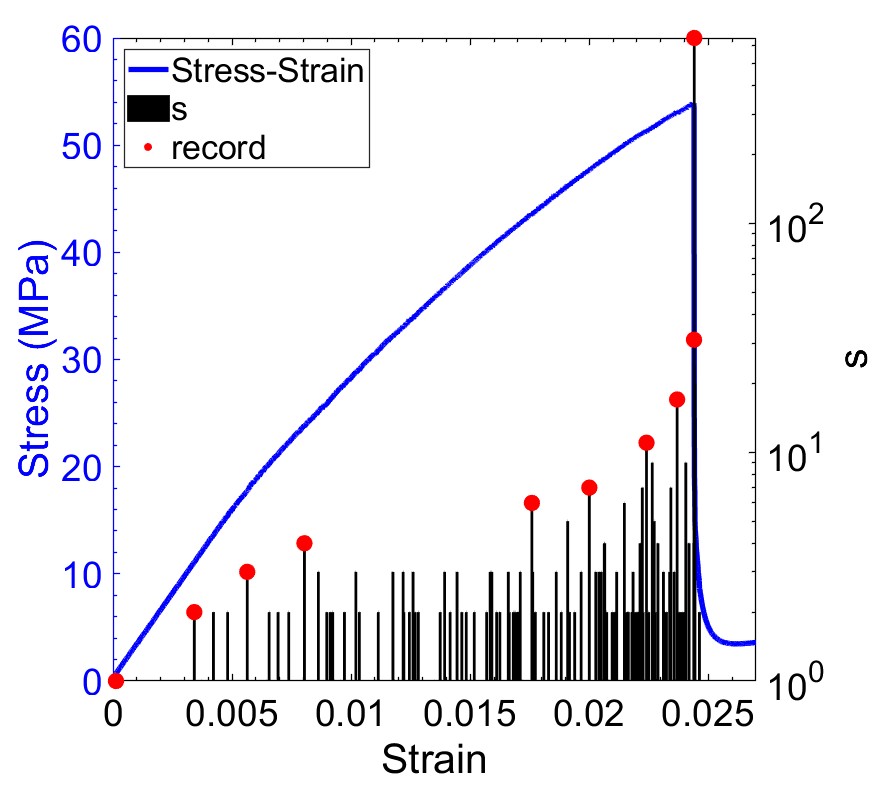}
\caption{}
\label{fig1d}
\end{subfigure}
\caption{Stress-strain curve of a typical realisation with its corresponding avalanches and records for (a) $\Delta$ = 0.05, (b) $\Delta$ = 0.15, (c) $\Delta$ = 0.2, and (d) $\Delta$ = 0.3.}
\label{fig1}
\end{figure}

\begin{figure}
\centering
\begin{subfigure}[b]{0.085\textwidth}
\centering
\textbf{$\Delta$ = 0.05}
\includegraphics[trim=0.0cm 0.0cm 0.0cm 1.0cm, clip,width=2 cm, height = 7.7 cm]{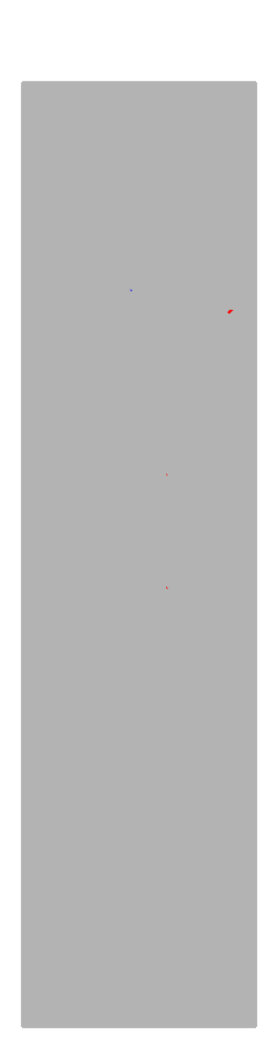}
\caption{}
\label{fig13a}
\end{subfigure}
\begin{subfigure}[b]{0.085\textwidth}
\centering
\textbf{$\Delta$ = 0.05}
\includegraphics[trim=0.0cm 0.0cm 0.0cm 1.0cm, clip,width=2 cm, height = 7.7 cm]{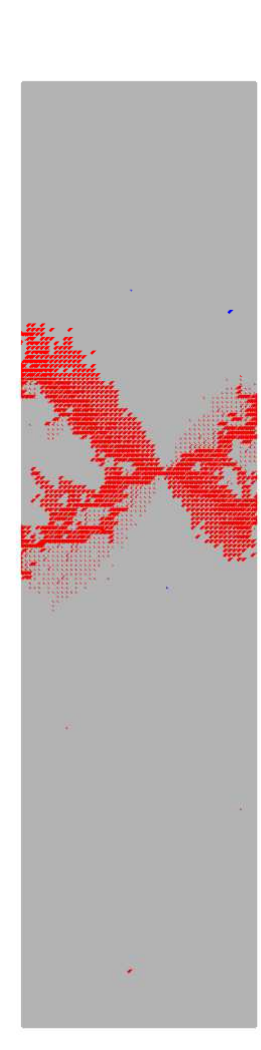}
\caption{}
\label{fig13b}
\end{subfigure}
\begin{subfigure}[b]{0.085\textwidth}
\centering
\textbf{$\Delta$ = 0.3}
\includegraphics[trim=0.0cm 0.0cm 0.0cm 1.0cm, clip,width=2 cm, height = 7.7 cm]{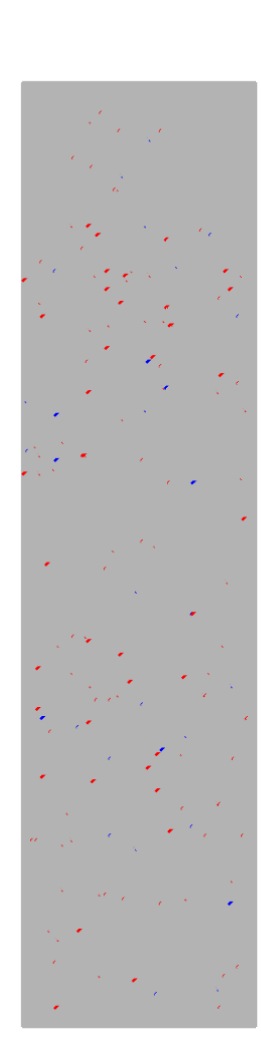}
\caption{}
\label{fig13c}
\end{subfigure}
\begin{subfigure}[b]{0.085\textwidth}
\centering
\textbf{$\Delta$ = 0.3}
\includegraphics[trim=0.0cm 0.0cm 0.0cm 1.0cm, clip,width=2 cm, height = 7.7 cm]{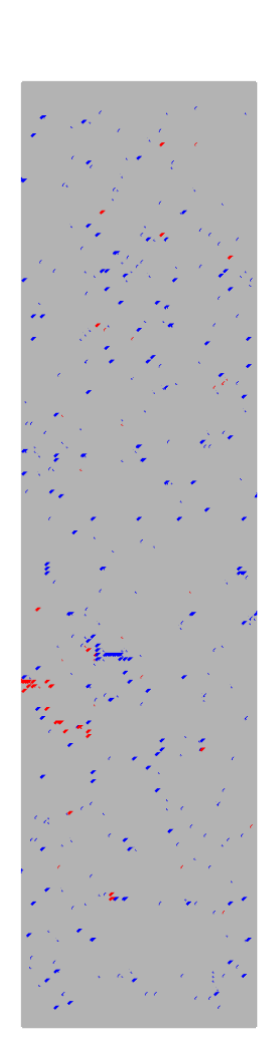}
\caption{}
\label{fig13d}
\end{subfigure}
\begin{subfigure}[b]{0.08\textwidth}
\centering
\textbf{$\Delta$ = 0.3}
\includegraphics[trim=0.0cm 0.0cm 0.0cm 1.0cm, clip,width=1.5 cm, height = 7.7 cm]{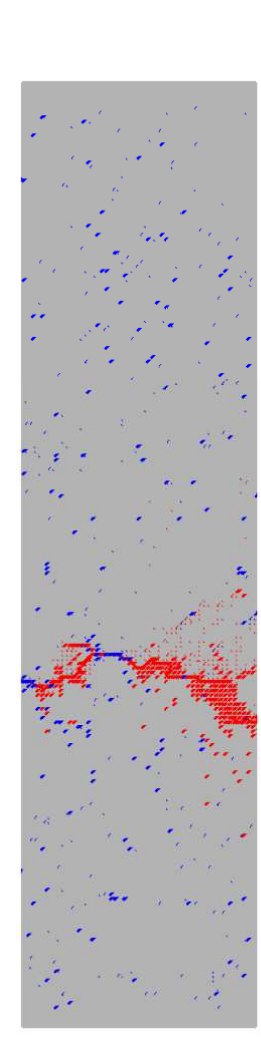}
\caption{}
\label{fig13e}
\end{subfigure}
\caption{Contour plots showing the evolution of fracture surfaces between consecutive records at different stages of loading. The grey, blue and red area represents the unbroken bonds at the end of the consecutive records, failed bonds before the consecutive records and the bonds fail between the consecutive records, respectively. (a) and (b) show the fracture surface evolved between [$1^{st}$, $2^{nd}$] and [$2^{nd}$, $3^{rd}$] records, respectively, for $\Delta$ = 0.05.  Fig (c), (d) and (e) show the fracture surface evolved between [$4^{th}$, $5^{th}$], [$8^{th}$, $9^{th}$] and [$9^{th}$, $10^{th}$]records, respectively, for $\Delta$ = 0.3. }
\label{fig13}
\end{figure}

To establish the sensitivity of the statistical signatures of the avalanche data to the RSNM model parameters, we first perform a parametric study. Initially, we look at the effect of disorder on the macroscopic response and the associated avalanche time series. The results for a range of $\Delta$ = 0.05, 0.15, 0.2 and 0.3, are shown in \cref{fig1}. The avalanche size, represented by $s$, is the number of springs broken for a given incremental strain. In the figure, we also highlight the record-breaking avalanches in red circles. A record-breaking avalanche is an avalanche whose size surpasses those of all previous avalanches. To study the avalanche and record up to the macroscopic failure strain of the material, we have chosen $\overline{\epsilon}_f$ = 0.03, 0.045, 0.057, and 0.074 for $\Delta$ = 0.05, 0.15, 0.2 and 0.3, respectively. The $\overline{\epsilon}_f$ are chosen such that for each $\Delta$, the stress-strain response averaged over 100 realisations closely reproduces the macroscopic failure strain. In spite of the randomness of the avalanche sizes and their time of occurrence, the macroscopic response is smooth for the chosen range of disorders. For low disorder, very few avalanches are observed and they are confined to the deformation regime close to final failure. With increasing disorder, the overall count of avalanches increases and avalanches are observed in several stages of deformation, including in the initial elastic regime for $\Delta$ greater than $0.2$. The record-breaking events for low disorder appear to occur rapidly till final failure. However, for the higher disorder, $\Delta$=$0.3$, we see a regime after the initial few records when there is a long gap before the next record-breaking event. In this stationary regime, the fracture process is dominated by the uncorrelated nucleation of micro-cracks, which are a consequence of the higher disorder. With further deformation as the fracture process approaches final failure, the subsequent record-breaking events occur more rapidly when the damage starts to develop longer-ranging correlations.

\begin{figure}
\centering
\begin{subfigure}[b]{0.237\textwidth}
\centering
\includegraphics[width=4.5cm]{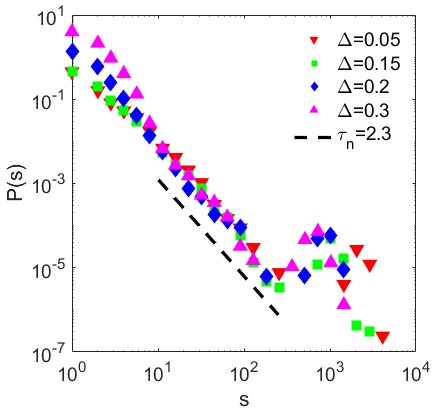}
\caption{}
\label{fig2a}
\end{subfigure}
\hfill
\begin{subfigure}[b]{0.237\textwidth}
\centering
\includegraphics[width=4.5cm]{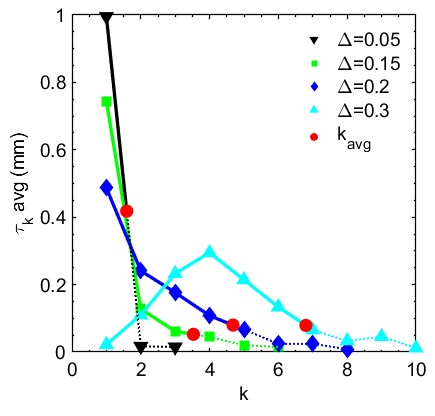}
\caption{}
\label{fig2b}
\end{subfigure}
\caption{(a) The probability of an avalanche of size $s$ and (b) the average waiting time of record avalanches for different $\Delta$. In (a), the data have been shifted vertically such that the data overlap for intermediate $s$.}
\label{fig2}
\end{figure}

To bring out the difference between the manner of damage propagation at different stages of deformation we look in detail at the spatial patterns of damage. For ease in distinguishing we colour code the springs that break in the given duration as red while the springs that were already broken before the duration are coloured blue. The remaining unbroken springs are coloured grey. At lower disorder ($\Delta$ = 0.05) the total number of records, for the typical realisation considered, is only $3$. In the duration between 1st and 2nd record very few springs break, appearing as isolated independent red regions, see \cref{fig13a}, while the rapid propagation of damage is seen between the 2nd and 3rd record, \cref{fig13b}. For higher disorder, which has distinct regimes in the waiting time intervals observed during the failure process, for the record that has the longest waiting time (between 4th and 5th record) in the contour plot in \cref{fig13c} the new events (red regions) are largely independent with very few in the neighbourhood of existing damage sites (blue regions), implying that in this duration the springs break throughout the domain like independent nucleation sites largely oblivious to the prior defect structure. In the waiting time for the last record, however, most of the red regions are localised as seen in \cref{fig13e} and several are part of a large cluster having connections with blue regions. This localisation of avalanches plausibly causes the waiting time between records to decrease. Once many springs are broken creating damage in different locations, it becomes easier to break the intermediate springs connecting the damaged areas, which in turn triggers more breakages, thus making it easier for larger avalanches to occur.

The role of disorder on different statistical signatures, the avalanche size distribution and the waiting time interval between record breaking events, are presented in \cref{fig2}. The probability of an avalanche of size $s$ is P(s) and its distribution follows a power law of the form: $P(s) \sim s^{-\tau_n}$ for $s>>1$. In \cref{fig2a}, to compare the slopes of P(s), the data for different disorders is shifted along the $y$-axis so that the data for intermediate $s$ overlap. For large avalanche sizes the power law exponent seems to be similar for all disorders and is approximately $2.3$. This suggests that for the range of disorder considered here, the effect of disorder on the exponent of avalanche size distribution is not sensitive
enough to characterise the disorder from the exponent. 

To understand the effect of disorder on the rate of occurrence of record breaking avalanches, we define the waiting time between two consecutive records $r_k$ and $r_{k-1}$ as: 
\begin{equation}
\label{eq12} 
\tau_k = t_{k} - t_{k-1},
\end{equation}
where $t_{k-1}$ and $t_{k}$ are the timestamp of the ${k-1}^{th}$ and ${k}^{th}$ record events. In \cref{fig2b} the variation of $\tau_k$ with $k$, averaged over 100 realisations, is shown for different disorder $\Delta$. The maximum of the range of $k$ indicates the highest number of records among all the realisations, and the red circles show the average of the number of records. From the record statistics of the simulation data, the waiting time between record breaking events at lower disorder ($\Delta$=$0.05$) is initially large and shows a rapid monotonic decrease with increasing k. In contrast, for higher disorder ($\Delta$=$0.3$) the waiting time is initially very short, it becomes longer with increasing $k$ upto $k=3$ where it peaks. With further increase in $k$, the waiting time reduces implying that as the system approaches final failure, record breaking events occur with shorter and shorter waiting time. While approaching to final failure acceleration in occurrence of record breaking events is seen in all cases of disorder, only high disorder exhibits the initial deceleration resulting in a peak. The peak shifts to the left as the disorder is reduced. The effect of disorder is also seen to affect the
overall count of record breaking events. For higher disorder the total number of record breaking events is higher.

\begin{figure}
\centering
\begin{subfigure}[b]{0.237\textwidth}
\centering
\includegraphics[width=4.5cm]{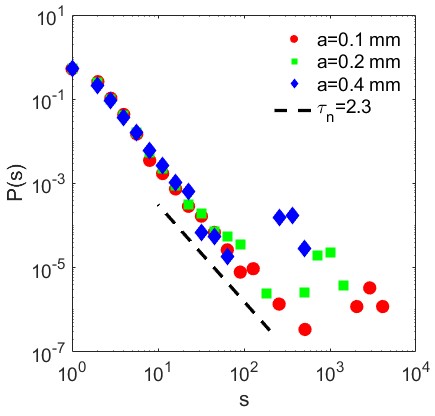}
\caption{}
\label{fig3a}
\end{subfigure}
\hfill
\begin{subfigure}[b]{0.237\textwidth}
\centering
\includegraphics[width=4.5cm]{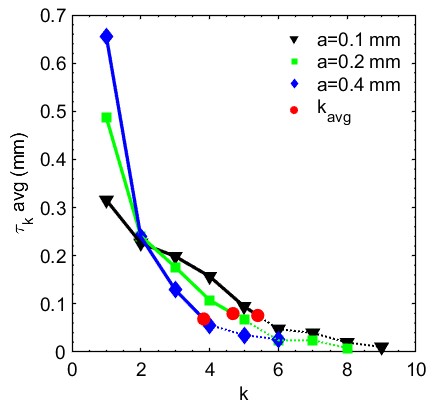}
\caption{}
\label{fig3b}
\end{subfigure}
\caption{(a) The probability of an avalanche of size $s$ and (b) the average waiting time of record avalanches for the $\Delta$ = 0.2 and different lattice sizes $a$.}
\label{fig3}
\end{figure}

In the RSNM simulations, lattice spacing is also an input parameter. To evaluate the sensitivity of the statistical signatures to the lattice spacing for a specific degree of disorder ($\Delta$ = 0.2), the lattice spacing of a= $0.1, 0.2$ and a=$0.4$ mm were chosen for the simulation. These lattice spacings fall within the acceptable range of RVE size suitable for modelling fracture in epoxy based polymer~\cite{lemaitre1986}. The avalanche distributions for all lattice spacings, as shown in \cref{fig3a}, overlap each other implying minimal effect on the power law exponent of the distribution. The waiting times however are affected by the lattice parameter $a$. Increasing system size (same as decreasing lattice spacing) seems to increase the overall number of records and the behaviour shows qualitative similarity to the effect of increasing disorder as seen in \cref{fig3b}. This implies that the effect of the system size and the disorder on waiting time interval is coupled and thus, the estimate of disorder parameter for a specific material system would be dependent on the choice of the system size. From the parametric study, we conclude that records are more sensitive to disorder than avalanche distribution, thus making records more suitable for characterising material disorder. 

\subsection{Experimental results}
\label{sec:4.2}

\begin{figure}
\centering
\begin{subfigure}[b]{0.237\textwidth}
\centering
\includegraphics[width=4.2cm]{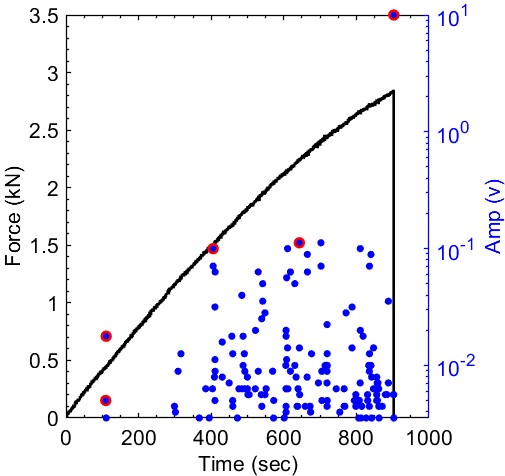}
\caption{}
\label{fig4a}
\end{subfigure}
\begin{subfigure}[b]{0.237\textwidth}
\centering
\includegraphics[width=4.2cm]{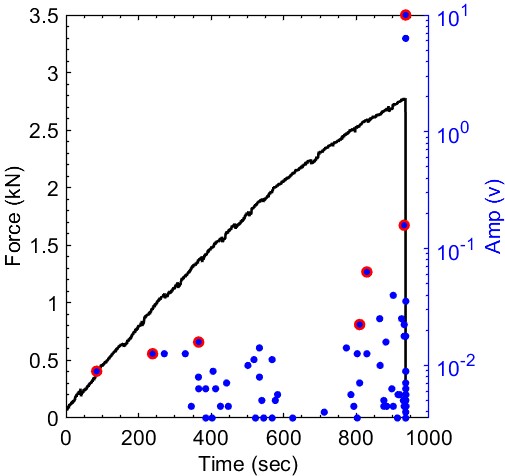}
\caption{}
\label{fig4b}
\end{subfigure}
\begin{subfigure}[b]{0.237\textwidth}
\centering
\includegraphics[width=4.2cm]{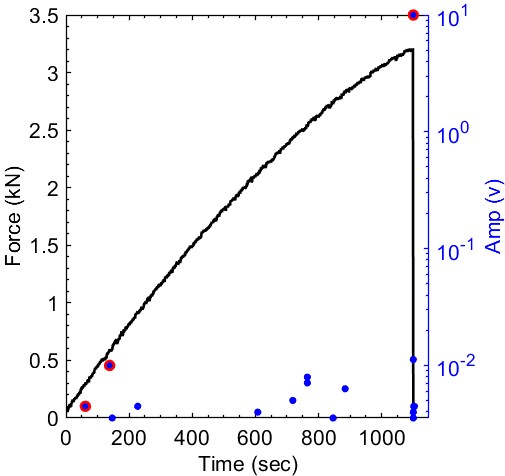}
\caption{}
\label{fig4c}
\end{subfigure}
\begin{subfigure}[b]{0.237\textwidth}
\centering
\includegraphics[width=4.2cm]{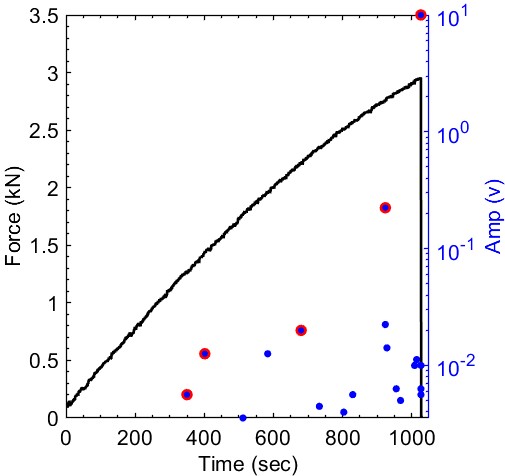}
\caption{}
\label{fig4d}
\end{subfigure}
\begin{subfigure}[b]{0.237\textwidth}
\centering
\includegraphics[width=4.2cm]{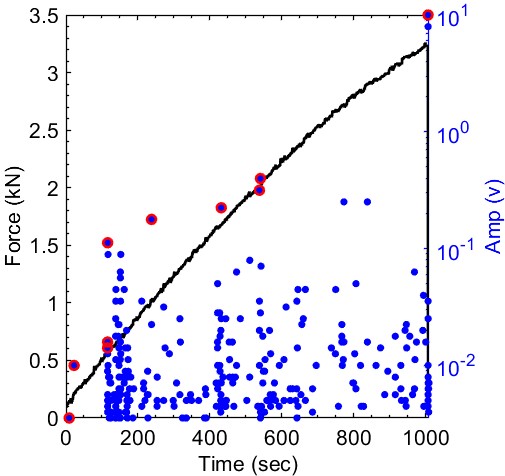}
\caption{}
\label{fig4e}
\end{subfigure}
\begin{subfigure}[b]{0.237\textwidth}
\centering
\includegraphics[width=4.2cm]{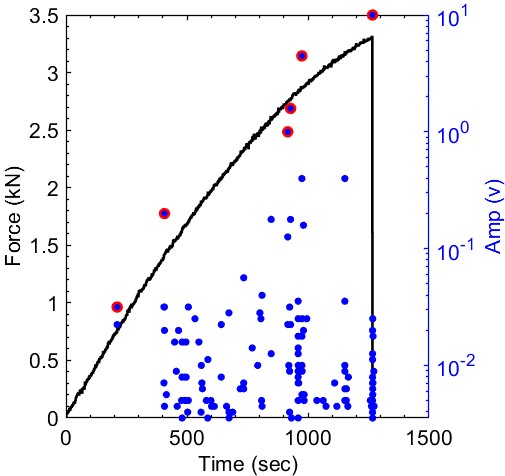}
\caption{}
\label{fig4f}
\end{subfigure}
\caption{Force-time response with its corresponding avalanches and record bursts from experiments, for specimens: (a) B1S1, (b) B1S3, (c) B3S1, (d) B3S3, (e) B4S2 and (f) B4S4. The blue and red circled blue dots correspond to avalanche and record avalanche, respectively.}
\label{fig4}
\end{figure}

\begin{figure}
\centering
\begin{subfigure}[b]{0.237\textwidth}
\centering
\includegraphics[width=4.2cm]{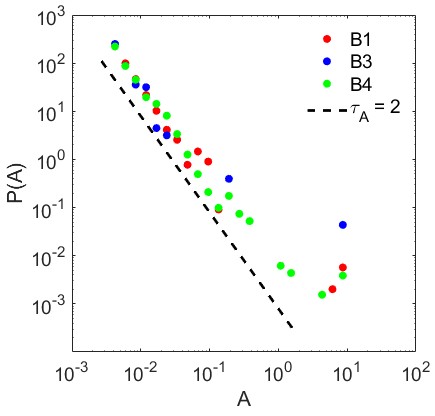}
\caption{}
\label{fig5a}
\end{subfigure}
\hfill
\begin{subfigure}[b]{0.237\textwidth}
\centering
\includegraphics[width=4.2cm]{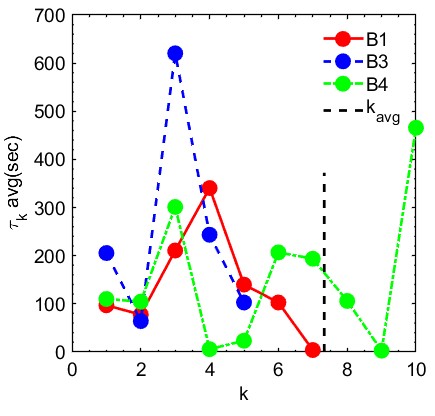}
\caption{}
\label{fig5b}
\end{subfigure}
\caption{(a) Avalanche distribution in Amp (volt) for three batches.(b) Average waiting time of records for three batches. [The x-intercept of the vertical dashed line indicates the average of the number of records ($k_{avg}$) detected in the specimens B1S1, B1S3, B3S1, B3S3, B4S2, and B4S4.]}
\label{fig5}
\end{figure}

For estimation of the disorder parameter in the model we examine the AE activity data for the equivalent statistical signatures. From the tensile test data of 6 samples prepared in three different batches (B1, B3, B4), the time-synced force (kN) and amplitude (volt) data, presented in \cref{fig4}, shows characteristics of quasi-brittle fracture, such as multiple events spread through the deformation to final failure which may be attributed to nucleation of uncorrelated micro-cracking in the early stages, followed by the growth and interactions of the micro-cracks represented by the medium amplitude AE events, and finally, the complete fracture of the specimen corresponding to the highest amplitude event that coincides with the final catastrophic load drop in the specimen. The quasi-brittle fracture signatures are also observed in the rate of record events (highlighted by the red circles). These AE signatures will be analysed next for characterising the degree of disorder. 

The avalanche distributions of the collective amplitudes recorded from the specimens of each batch are presented in \cref{fig5a}. The avalanche distributions of all the batches are very similar to each other and follow a power law distribution: $P(A) \sim A^{-\tau_A}$ with $\tau_A\approx2$. The waiting time for the records, obtained from averaging only the non-zero entries for each record, are presented in \cref{fig5b}. All the batches exhibit a peak (at 3rd or 4th record). The batch B4, however, shows anomalous behaviour with two additional peaks. The anomaly is due to the specimen B4S2 exhibiting large cluster of AE activity in the initial stages (at time ranging between 100-200 secs) which is not observed in any other specimen. In further analysis, the experimental data from B4S2 is ignored.

\subsection{Experimental validation}
\label{sec:4.3}

\begin{figure}
\centering
\includegraphics[width=5.5cm]{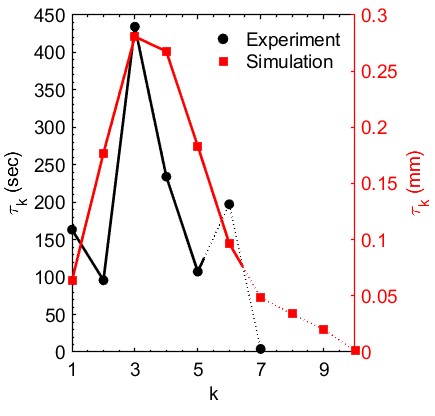}
\caption{The variation of waiting time averages for the experiment record bursts are compared with the corresponding records from simulation for $\Delta$ = $0.275$.}
\label{fig8}
\end{figure}

It is understood from the parametric study that the waiting time interval and the total number of records have significant sensitivity to the disorder incorporated in the model. From the experimental data of all batches (excluding B4S2), we find the average waiting time for every record by considering only the nonzero $\tau_k$ (see \cref{fig8}). A distinct peak is observed at $k=3$, and the averaged total number of records to be between 5 and 6. Comparison with the simulation results in \cref{fig2b} suggests the $\Delta$ to fall between $0.2$ to $0.3$. Iteratively the combination of  $\overline{\epsilon}_f$ = 0.0715 and $\Delta$ = 0.275 is found to closely reproduce the features of the experimental data, as seen in \cref{fig8}. The lattice spacing is chosen to be $\mathrm{a=0.2 mm}$, same as the initial parametric study.

\begin{figure}
\centering
\begin{subfigure}[b]{0.237\textwidth}
\centering
\includegraphics[width=4.2cm]{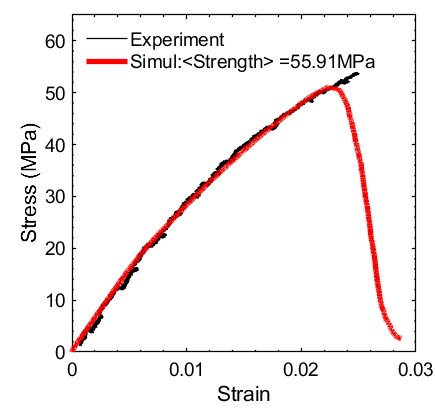}
\caption{}
\label{fig10a}
\end{subfigure}
\hfill
\begin{subfigure}[b]{0.237\textwidth}
\centering
\includegraphics[width=4.2cm]{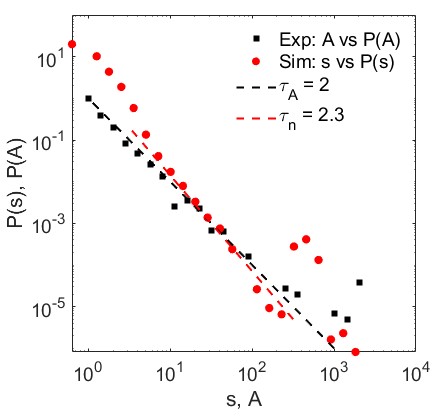}
\caption{}
\label{fig10b}
\end{subfigure}
\caption{(a) The stress-strain response of the B1S3 specimen from the experiment is compared with the averaged stress-strain response from the simulation for $\Delta$ = 0.275,  and (b) their corresponding avalanche distributions. In the avalanche distribution, the unit of $s$, $A$, $P(s)$ and $P(A)$ is arbitrary and for coinciding the middle portion of the graphs, the graphs are shifted by maintaining the same slope.}
\label{fig10}
\end{figure}

\begin{figure}
\centering
\begin{subfigure}[b]{0.49\textwidth}
\centering
\includegraphics[width=6.0cm]{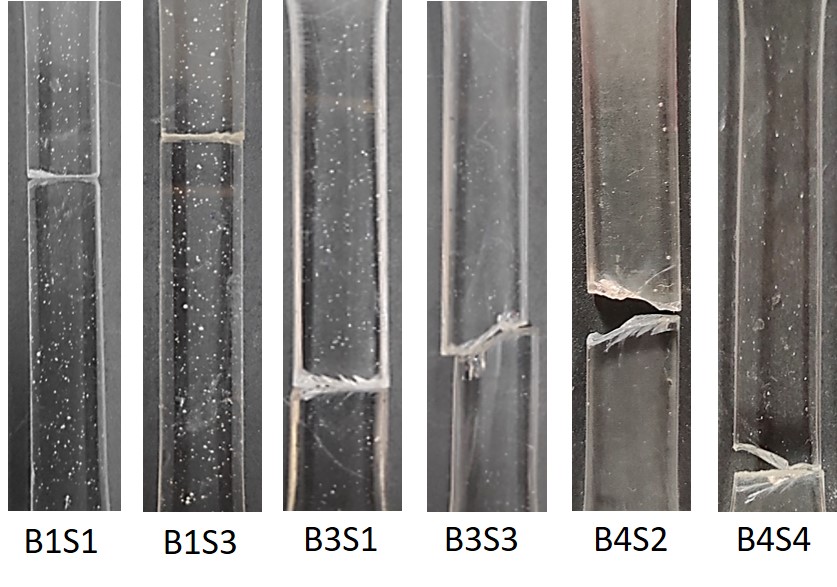}
\caption{}
\label{fig11a}
\end{subfigure}
\hfill
\begin{subfigure}[b]{0.49\textwidth}
\centering
\includegraphics[width=5.7cm, height=4.275cm]{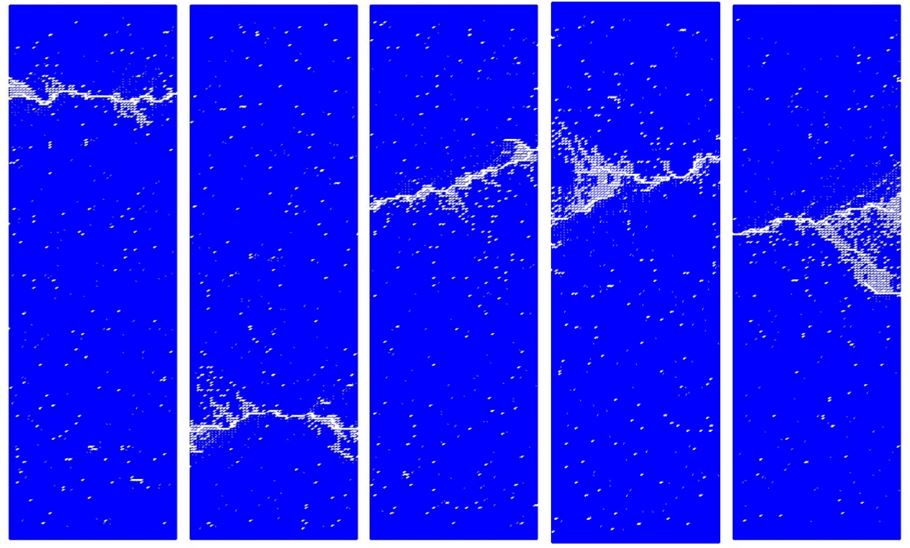}
\caption{}
\label{fig11b}
\end{subfigure}
\caption{(a) The fracture surface in the gauge section of the failed specimens and (b) the simulated fracture path for different realisations.}
\label{fig11}
\end{figure}

\begin{figure}
\centering
\begin{subfigure}[b]{0.237\textwidth}
\centering
\includegraphics[width=4.5cm]{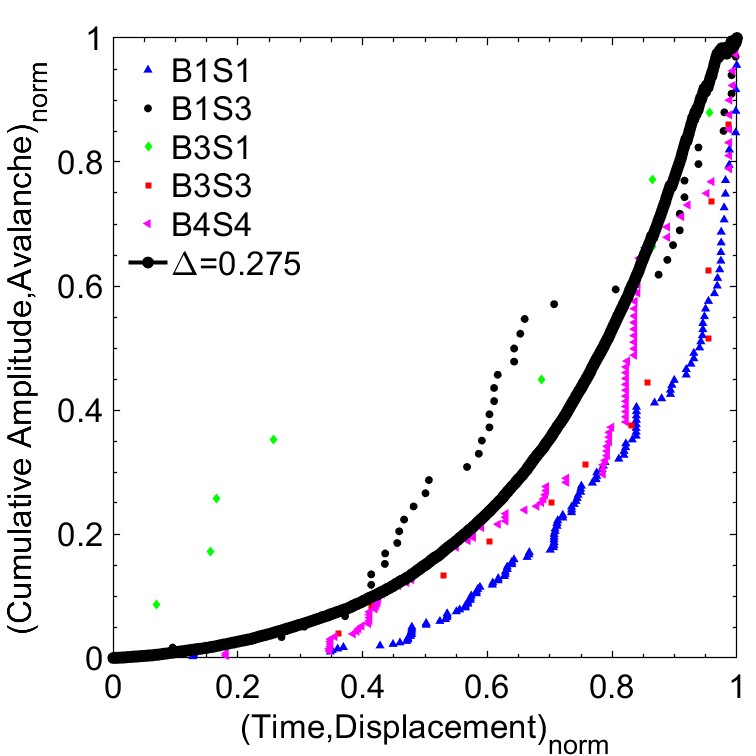}
\caption{}
\label{fig12a}
\end{subfigure}
\hfill
\begin{subfigure}[b]{0.237\textwidth}
\centering
\includegraphics[width=4.5cm]{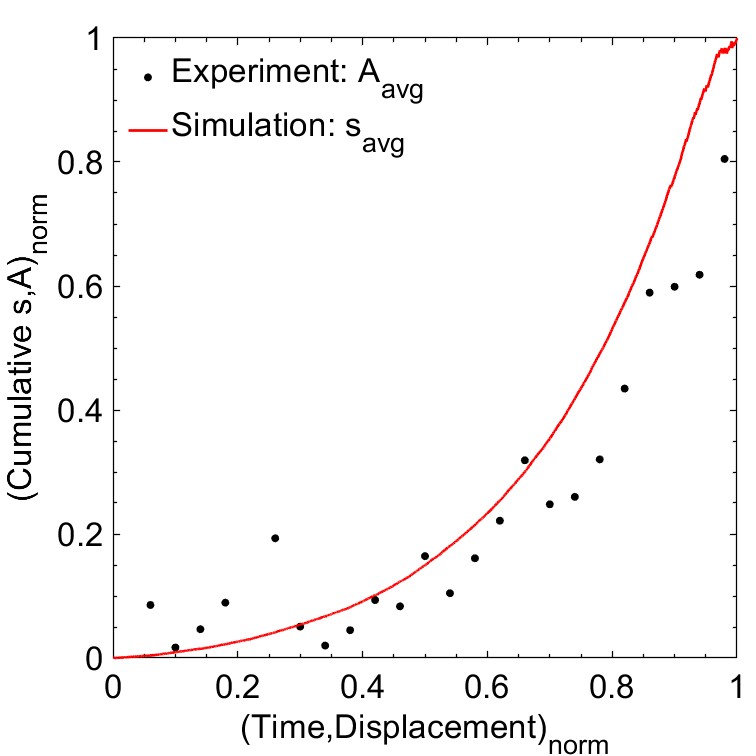}
\caption{}
\label{fig12b}
\end{subfigure}
\caption{(a) In the time or displacement vs. the cumulative damage growth graph, first, both the x and y data are normalised with their respective maximums, (b) then the normalised avalanches are averaged over the realisations for the simulation and the normalised amplitudes are averaged over the samples in the experiment}
\label{fig12}
\end{figure}

The effectiveness of the chosen model parameters is further examined by comparing the simulation data with other features of the experiments. The macroscopic stress-strain response averaged over 250 realisations, shown in \cref{fig10a}, well reproduces experimental curve in its linear and non-linear regime. The averaged strength from simulations is within $2\%$ of the average experimental strength. The avalanche size distribution, P(s), and the AE amplitude distribution, P(A), both exhibit power law behaviour in the large avalanche size regime, the exponents being 2 and  2.3. 

Further, we examine the ability of the model to simulate realistic fracture paths. In \cref{fig11a} we note that several types of fracture paths are seen in experiments, nearly flat (B1S1, B1S3, B3S1), tilted (B3S3), branched (B2S2, B4S4). Such features are well reproduced in a few selected realisations from the simulations in \cref{fig11b}. The model, thus, for the estimated disorder parameter is able to simulate close to realistic fracture paths. 

In \cref{fig12}, over the course of deformation to final failure the progressive growth of damage is evident in the cumulative amplitude and cumulative avalanche size from experiments and simulations respectively. The cumulative data is normalised with the final value for comparison of the rates. The comparison of averaged experimental data and simulations, shown in \cref{fig12b}, further confirms that incorporating a suitable degree of disorder in RSNM can adequately simulate the damage nucleation and growth activity in glassy epoxy.

\section{Conclusion}
\label{sec:5}

In the present work, we show that the statistical signatures of AE activity during the failure process can be utilised for characterisation of disorder parameter in simulation of tensile fracture of epoxy based polymer. For simulations we use a square random spring network model with nearest and next-nearest interactions. The spring behaviour is taken to account for quasi-brittle behaviour with linear and non-linear regimes and a normally distributed failure strain threshold. In the initial parametric study, we show that the disorder characteristics while have no noticeable effect on the power law exponent of the avalanche size distribution, are strongly correlated with the waiting time interval between consecutive record breaking avalanches as well as the total number of records. This sensitivity to disorder  is exploited in estimating the disorder parameter suitable for experiments on tensile failure of epoxy based polymer. Assuming equivalence between the amplitude distribution of AE data and avalanche size distribution of the simulations, we choose the $\Delta$ to be $0.275$ on the basis that the waiting time interval characteristics and total number of record breaking events are seen to correlate well between experiments and simulations for it. The chosen disorder parameter is next shown to well reproduce the failure characteristics in terms of the peak load of the macroscopic response, the power-law behaviour with avalanche dominated fracture type as well as realistic fracture paths. Utilising AE data for characterising disorder has enormous potential in fracture modelling of brittle, heterogeneous material systems as it can overcome one of the major drawbacks of local approach to fracture modelling: difficulty in estimation of model parameters from direct experimental data.

 \bibliographystyle{elsarticle-num} 
 \bibliography{vManuscript}





\end{document}